\documentclass	[english,aps,prb,reprint, superscriptaddress]{revtex4-2}
\usepackage{bm}
\usepackage[T1]{fontenc}
\usepackage[latin9]{inputenc}
\usepackage{xcolor}
\usepackage{float}
\usepackage{amsmath}
\usepackage{amssymb}
\usepackage{graphicx}

\makeatletter

\usepackage{grffile}
\usepackage{amsmath}

\renewcommand{\fnum@figure}{FIG. \thefigure}

\makeatother

\usepackage{babel}

\begin{document}

\author{Yonatan Messica}
\affiliation{
	Department of Physics, Bar-Ilan University, Ramat Gan, 52900, Israel
}
\author{Dmitri B. Gutman}
\affiliation{
	Department of Physics, Bar-Ilan University, Ramat Gan, 52900, Israel
}
\author{Pavel M. Ostrovsky}
\affiliation{
	Max Planck Institute for Solid State Research, Heisenbergstrasse 1, 70569, Stuttgart, Germany
}
\affiliation{
	L. D. Landau Institute for Theoretical Physics RAS, 142432 Chernogolovka, Russia
}

\begin{abstract}
We study the anomalous Hall effect in a disordered Weyl semimetal.
While the intrinsic contribution is expressed solely in terms of Berry curvature, 
the extrinsic contribution is given by a combination of the skew scattering and
side jump terms.
For the model of small size impurities, we are able to express the skew scattering contribution in terms of scattering phase shifts.  We identify the regime in which the skew scattering contribution dominates the side-jump contribution: the impurities are either strong or resonant, and at dilute concentration. In this regime, the Hall resistivity $\rho_{xy}$ is expressed in terms of
two scattering phases, analogous to the s-wave scattering phase in a non-topological metal.
We compute the dependence of $\rho_{xy}$ on the chemical potential, and show that $\rho_{xy}$ scales with temperature as $T^2$ in low temperatures and as $T^{3/2}$ in the high temperature limit.
\end{abstract}

\title{Anomalous Hall effect in disordered Weyl semimetals}
\maketitle

\section{Introduction}

Although the anomalous Hall effect (AHE) has been discovered experimentally
in 1881 \cite{Hall1881},  understanding its microscopic origins took longer to come, and its connection to topological properties of the electronic band structure has only been recently realized \cite{Nagaosa2010,Sinitsyn2007}. 

Much of the attention has been focused on materials with strong spin-orbit interaction.
In this case, several mechanisms contribute to the AHE. These mechanisms can be divided into intrinsic and extrinsic mechanisms.
The intrinsic contribution depends solely on the band structure of the material, and its origin is the anomalous velocity of the electrons due to their Berry curvature. The extrinsic contribution involves scattering of electrons, and semiclassicaly it can be separated to the skew-scattering and side-jump mechanisms \cite{Sinitsyn2006, Sinitsyn2007, Xiao2010, Ado2016}.
These competing mechanisms give rise to the complexity of the AHE. The magnitudes of the different processes scale differently with carrier density and the nature of the disorder, making the behavior of the AHE rich and complex.


In this work we focus on the problem of AHE in Weyl semimetals (WSMs) with broken time-reversal symmetry (TRS).
WSMs have a strong spin-orbit interaction necessary for the AHE, and moreover the Weyl nodes are sources of Berry curvature \cite{Jia2016, Armitage2018, Yan2017}.
When the chemical potential is in the vicinity of the Weyl nodes, the intrinsic mechanism dominates the AHE, giving rise to the pseudo-quantized AHE conductivity \cite{Burkov2011}. 
However, away from that point, Fermi surfaces are formed around Weyl nodes and the extrinsic mechanisms contribute as well. The relative role of intrinsic and extrinsic processes for different spectrum was studied experimentally for the WSMs PrAlGe$_{1-x}$Si$_x$ and Co$_3$Sn$_2$S \cite{Yang2020, Shen2020}. By continuously varying the doping, the relative importance of the extrinsic and intrinsic mechanisms was identified using scaling analysis \cite{Shitade2012, Hou2015}.
To understand such phenomena, one needs to construct a model containing the essential ingredients of these materials that allows an analytic solution.
This is what we do in this current work.
We will focus on a disorder regime in which the calculation of the AHE resistivity is greatly simplified, as it is dominated by the extrinsic skew-scattering contribution. We derive a simple formula for $\rho_{xy}$, expressing it in terms of scattering phases, with which experimental results can be readily compared. Let us mention that recent works have analyzed the problem of anomalous Hall conductivity in tilted WSMs \cite{papaj2021enhanced, zhang2023disorder, zhang2023anomalous}. In these works, a finite Born approximation was taken to consider the impurity scattering. Contrarily, in the present work, we will consider impurities of arbitrary strength, where the solution of the full scattering problem is necessary.
We also emphasize that we focus on the regime of weak disorder, where the disorder affects the transport properties but not the band structure. In the regime of strong disorder, a phase transition is expected from a WSM to a diffusive metal through a Chern insulator state \cite{Chen2015, Liu2016, Su2017}. The subtle effect of resonant scattering inducing a finite density of states at the Weyl points \cite{Pixley2021, Holder2017} is also beyond the scope of our work.

\section{Model}

To study the AHE in 3D Weyl semimetals, we consider the following
minimal model for a single pair of Weyl nodes:

\begin{equation}
\label{eq:Hamiltonian}
H=u_{\Vert}\left(p_{x}\sigma_{x}+p_{y}\sigma_{y}\right)+\left(\frac{p_{z}^{2}}{2m}-\lambda\right)\sigma_{z}+V.
\end{equation}
Here $\sigma_{i}$ are Pauli matrices in the pseudo-spin space and
$V$ is the static disorder potential due to impurities put at dilute concentration $n_i$. The $\sigma_{z}$
term breaks the time-reversal symmetry. The energy parameter $\lambda$ controls the spacing between the Weyl nodes. We will study the Hall response
in the xy-plane, in which the model has rotation symmetry.

\begin{figure}[t]
	\centering
	\includegraphics[scale=0.8]{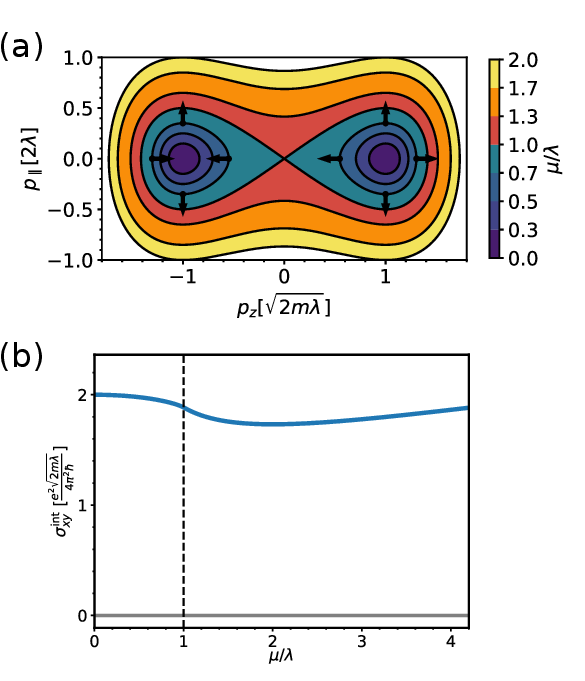}
	\caption{\label{fig:fermi surface and sigma_xy intrinsic} (a) Fermi surfaces for the WSM Hamiltonian. $p_\parallel$ is the projection of $\bf{p}$ to the x-y plane. For $\mu<\lambda$, there are two surfaces surrounding each Weyl node, which merge at $\mu=\lambda$. The arrows indicate the direction of the spinor part of the eigenstates at the corresponding momentum points. (b) Intrinsic part of the AHE conductivity, $\sigma_{xy}^{\rm{int}}$.}
\end{figure}

The periodic part of the plane-wave eigenstates is a two-component spinor. The direction of the spinor is dependent on the momentum, giving rise to the rich dynamics as the electrons move in momentum space due to an external field or disorder scattering. The two Weyl nodes are located at $\mathbf{p}=(0,0,\pm \sqrt{2m\lambda})$. At chemical potential $\mu=0$, the Fermi surface consists only of the two Weyl points and the density of states vanishes. As the chemical potential is raised, two Fermi surfaces emerge around the Weyl nodes. At $\mu=\lambda$, the two surfaces merge through Lifshitz transition (Fig. \ref{fig:fermi surface and sigma_xy intrinsic}).

The transport properties of a material can be calculated semiclassically,
using the Boltzmann formalism. Generally, a modified Boltzmann equation is required in materials with non-trivial topology, to account for virtual processes involving several bands \cite{Sinitsyn2007}. However, in the regime of rare strong impurities considered in this work, it will suffice to consider the simple Boltzmann equation \cite{Ashcroft1976}:
\begin{equation}
\frac{\partial f_l}{\partial t}+e{\bf E}{\bf \nabla}_{\bf k} f_l = -\sum_{l'}w_{ll'}\left[f_l - f_{l'} \right].
\label{eq:Boltzmann general}
\end{equation}
Here, $w_{l'l}$ is the scattering rate from state $l'=\left|u_{b' \mathbf{k'}} \right\rangle$ to state $l=\left|u_{b \mathbf{k}} \right\rangle$.
The velocity operator is given by
\begin{equation}
\mathbf{v}_{l}=\mathbf{\nabla_k}\epsilon_{l}+\frac{d\mathbf{p}}{dt}\times\bm{\mathcal{F}}_{l}.
\end{equation}

In addition to the usual dispersion term, the Berry curvature of the band, defined by
$\bm{\mathcal{F}}_{l}=-i\left\langle \mathbf{\nabla}_{\mathbf{k}}u_{l}\right|\times\left|\mathbf{\nabla}_{\mathbf{k}}u_{l}\right\rangle $, 
gives rise to the second term known as the anomalous velocity \cite{Xiao2010}. The omission of the side-jump part of the velocity \cite{Sinitsyn2007} will be justified later on.

\begin{figure}[t]
\centering
\includegraphics[scale=0.5]{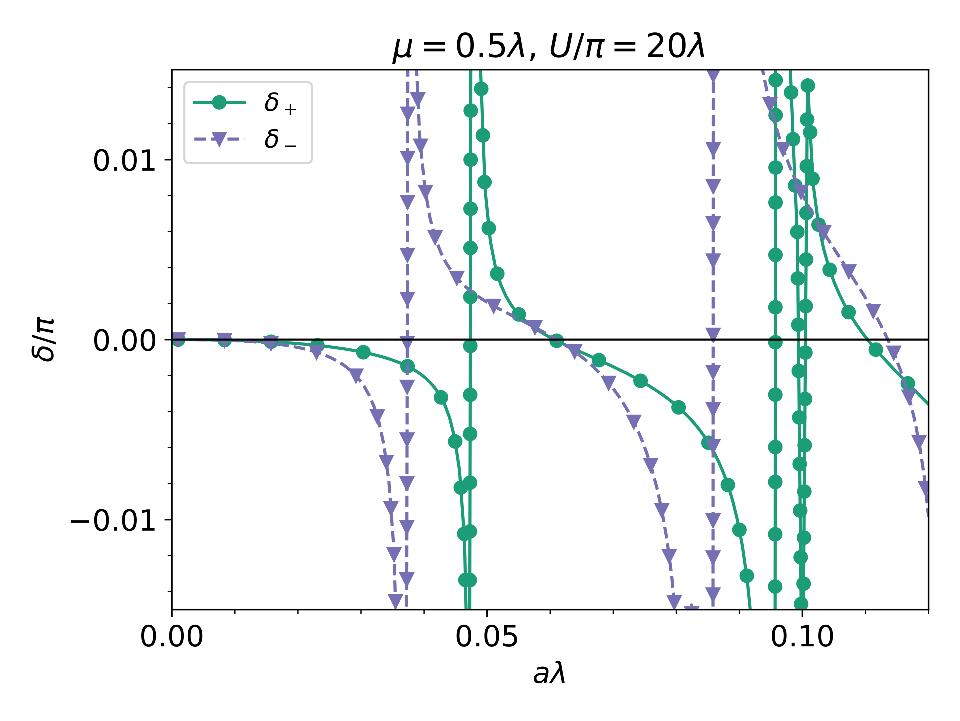}

\caption{\label{fig:scattering phases} The main scattering phases $\delta_{\pm}$ for varying impurity radius $a$, at a fixed chemical potential $\mu=0.5\lambda$. The strength
of the impurity potential $U=20\lambda$ and the aspect ratio $F=1$ are kept fixed. When the impurity radius $a$ goes to zero,
the scattering phases scale as $a^{5/2}$, while for $Ua\gtrsim 1$ they scale as $a^{3/2}$. Resonance occurs at small regions, which are different for $\delta_+$ and $\delta_-$.}
\end{figure}

To employ the Boltzmann equation one needs to compute the scattering rate $w_{ll'}$ by solving 
the scattering problem for the anisotropic spectrum [Eq. (\ref{eq:Hamiltonian})].
To account for the skew-scattering contribution of the anomalous Hall conductivity, one must go beyond the first
Born approximation in the calculation. However, solving the scattering problem analytically is
 non-trivial for an anisotropic model such as the one considered. Nonetheless, in the limit of point-like impurities (much smaller than the Fermi wavelength) we are able to solve the problem analytically, and thus we focus on this case from hereon.
One can show (Appendix \ref{sec:scat-phase-appendix}) that for a two-band model, the entire scattering problem is dominated
by two scattering phases $\delta_{\pm}$ for two scattering eigenstates.
This generalizes the usual electron s-wave scattering to Weyl fermions. The asymmetry between the plus and minus phases is due to the breaking of the time-reversal symmetry.
For point-like impurities, $\delta_{\pm}$ are typically small, except for small regions of resonance (Fig. \ref{fig:scattering phases}).



From the scattering phases we calculate the disorder scattering rate $w_{ll'}$. The asymmetry between $\delta_+$ and $\delta_-$ yields an asymmetric part to the scattering rate, such that $w_{ll'} \neq w_{l'l}$, giving rise to the skew-scattering extrinsic AHE.

After distribution function is obtained by the Boltzmann equation (\ref{eq:Boltzmann general}), the total current is computed by
\begin{equation}
{\bf j}=e\sum_{l}f_{l}{\bf v}_{l}.
\end{equation}

We decompose the anomalous Hall conductivity into intrinsic and extrinsic parts.
The former is independent of the disorder strength, and is determined by the integrated Berry curvature over the filled Fermi sea \cite{Sundaram1999, Haldane2004}:

\begin{equation}
\sigma_{xy}^{\rm int} =
e^2 \sum_{l}f_{l} {\left( \mathcal{F}_{l} \right)} _z. \label{eq:sigma_xy_int}
\end{equation}
The extrinsic contribution is due to the linear correction of the distribution function, which depends on the given disorder potential. Generally, it is made of skew-scattering and side-jump contributions \cite{Sinitsyn2007, Ado2016}, making the problem complicated. However, we will identify and focus on a regime where skew-scattering dominates.


\section{Results}

\subsection{Intrinsic contribution}
Now we proceed to calculate the intrinsic contribution. First, the Hamiltonian (\ref{eq:Hamiltonian}) needs to be regularized. This is because the low-energy description does not determine the location of the Fermi arcs; they can either connect the Weyl nodes from between through $\left| k_z \right| < \sqrt{2m\lambda}$, or away from them, at $\left| k_z \right|$ going to infinity. To account for the former case, we regularize the mass term multiplying $\sigma_z$ by changing $p_z^2/(2m) - \lambda \rightarrow \left(p_x^2 + p_y^2 + p_z^2\right)/(2m) - \lambda$. In this way, for a plane with fixed $p_z=q$, the mass either changes sign from $p_\parallel\equiv\sqrt{p_x^2+p_y^2}=0$ to $p_\parallel=\infty$ when q is between the Weyl nodes ($\lvert q \rvert < \sqrt{2m\lambda}$), or it remains with the same sign for $\lvert q \rvert > \sqrt{2m\lambda}$, giving rise to the Fermi arcs in the first region. We then compute the intrinsic contribution using Eq. (\ref{eq:sigma_xy_int}) and obtain the result depicted in Fig. \ref{fig:fermi surface and sigma_xy intrinsic}:

\begin{widetext}
\begin{equation}
\sigma_{xy}^{\mathrm{int}}(\epsilon)=\frac{e^{2}\sqrt{2m\lambda}}{2\pi h}\begin{cases}
\sqrt{1+\frac{\left| \epsilon \right|}{\lambda}}+\sqrt{1-\frac{\left| \epsilon \right|}{\lambda}} + \frac{\lambda}{3\left| \epsilon \right|} \left(\sqrt{1+\frac{\left| \epsilon \right|}{\lambda}}-\sqrt{1-\frac{\left| \epsilon \right|}{\lambda}} \right)
\left(1-\sqrt{1-\frac{\left| \epsilon \right|^2}{\lambda^2}}\right) & \left| \epsilon \right|<\lambda, \\
\frac{2}{3} \sqrt{1+\frac{\left| \epsilon \right|}{\lambda}} \left(1 + \frac{\lambda}{\left| \epsilon \right|} \right) & \left| \epsilon \right|>\lambda.
\end{cases}\label{eq:sigma_xy_int}
\end{equation}
\end{widetext}

Near the neutrality point we get the standard result \cite{Burkov2014}, $\sigma_{xy}^{\mathrm{int}}(\mu=0)=\frac{e^{2}}{2\pi h} \Delta_k $, where $\Delta_k =2 \sqrt{2m\lambda}$ is the distance between the Weyl nodes. Each filled 2D band with $p_z=q$ between the Weyl nodes contributes a to the Hall conductivity by the quantized value  $\sigma_{xy}^{\mathrm{2D}}=\frac{e^{2}}{h}$.
Moving the chemical potential away from the neutrality point changes the value of the intrinsic conductivity slowly as two Fermi surfaces are formed. A signature can be seen of the Lifshitz transition point at $\mu=\lambda$, where the two Fermi-surfaces merge. At this point, 
$\sigma_{xy}^{\rm int}(\mu=\lambda)\simeq 0.94 \sigma_{xy}^{\rm int}(\mu=0)$.

\subsection{Extrinsic contribution}

Next we discuss the extrinsic contribution due to scattering by the disorder potential. The extrinsic Hall conductivity will contain a term due to skew-scattering by a single impurity, which scales as $\sigma_{xy}^{\rm ext, 1} \propto n_i^{-1}$, and additional terms which are independent of the impurity concentration and originate from skew-scattering side-jump processes involving multiple bands \cite{Sinitsyn2007, Ado2016, papaj2021enhanced, zhang2023disorder, zhang2023anomalous}, which we label as $\sigma_{xy}^{\rm ext, 0}$. The following argument shows that for samples with sufficiently dilute concentration of impurities, the single impurity skew-scattering contribution $\sigma_{xy}^{\rm ext, 1}$ will dominate. From dimensional arguments, the terms must scale as $\sigma_{xy}^{\rm ext, 1}=g_1(\mu/\lambda) \frac{ {\left(m \lambda \right)}^2} {n_i \delta} $ and $\sigma_{xy}^{\rm ext, 0}=g_0(\mu/\lambda) \sqrt{m \lambda}$, where $g_1, g_0$ are some dimensionless functions which go to zero at $\mu=0$.
Therefore, at low enough impurity concentration such that $n_i \delta \ll (m\lambda)^{3/2}$, the single impurity skew-scattering contribution $\sigma_{xy}^{\rm ext, 1}$ always dominates $\sigma_{xy}^{\rm ext, 0}$.

This is the simplification that enables us to analytically solve the problem. The anomalous part of the collision integral, the side-jump part of the velocity operator and the off-shell skew-scattering processes all contain an extra factor of $n_i$ \cite{Sinitsyn2007}, giving a contribution to $\sigma_{xy}$ in which the dependence on $n_i$ cancels, and therefore they are negligible in our model.

Following the semiclassical Boltzmann formalism, we calculate the elastic scattering time $\tau^{\rm{el}}$ as well as the parallel and perpendicular transport times $\tau^{\parallel},\tau^{\perp}$
in our model (see Supplemental Material). Skew-scattering is enabled by asymmetry in the scattering phases of the two principal scattering modes. Parameterically, the inverse skew-scattering rate $1/\tau^{\perp}$ scales as $\sin(\delta_+ -\delta_-) / \tau^{\rm{el}}$, and is thus enhanced by stronger impurities.

The total Hall conductivity is given by the sum of the intrinsic and extrinsic contributions,
\begin{equation}
\sigma_{xy}\simeq \sigma_{xy}^{\rm int}+\sigma_{xy}^{\rm ext,1} \label{eq:sigma_xy_sum}.
\end{equation}
In the vicinity of the Weyl nodes, the extrinsic contribution vanishes, 
and the conductivity is dominated by the intrinsic term. Increasing the chemical potential, eventually there is a transition to the skew-scattered dominate regime. This is because the intrinsic contribution scales similarly to $\sigma_{xy}^{\rm ext, 0}$ at finite chemical potential, which we already have established is smaller than $\sigma_{xy}^{\rm ext, 1}$. The Hall resistivity is obtained by inverting the conductivity tensor. In the skew-scattering dominated regime, it is given by (denoting $\nu_{\pm}$ for the density of states projected on the up/down spinors in the z-axis)
\begin{equation}
\rho_{xy}=
-\frac{n_{i}\sin(\delta_{+})\sin(\delta_{-})}{\pi e^{2}\nu_{+}\nu_{-}}\sin\left(\delta_{+}-\delta_{-}\right).\label{eq:rho_xy two phases}
\end{equation}
In the vicinity of the Weyl node, The Hall resistivity is simply $\rho_{xy}=\rho_{xx}^{2}\sigma_{xy}^{\rm int}$, where $\rho_{xx}$ is the longitudinal resistivity (see Supplemental Material for explicit expressions).

Eq. (\ref{eq:rho_xy two phases}) is the main result of our work. The
Hall resistivity in the regime of strong, rare impurities is determined by two scattering
phases characterizing the impurities.

\begin{figure}[t]
\centering

\includegraphics[scale=0.42]{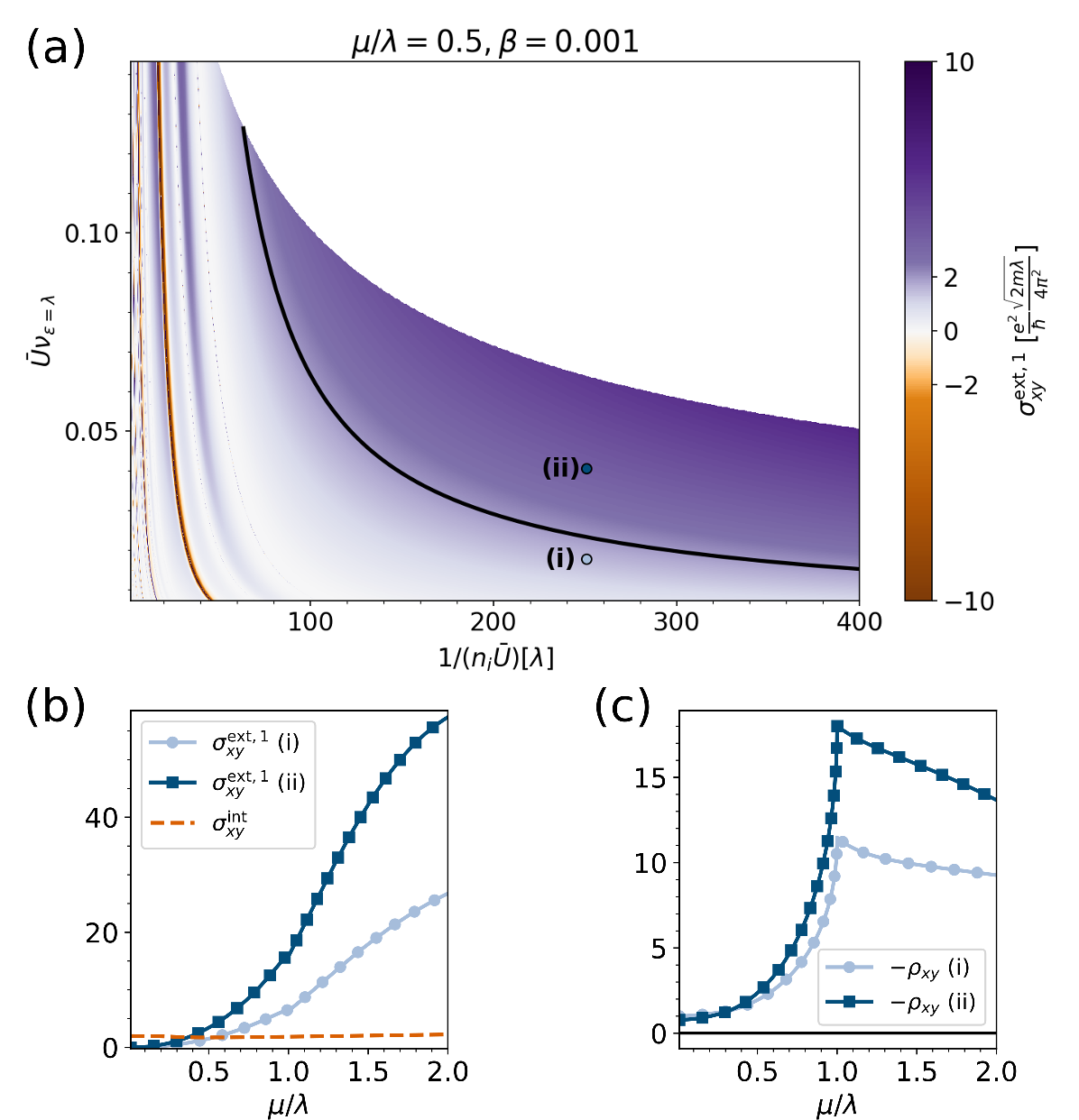}

\caption{(a) $\sigma_{xy}^{\rm{ext,1}}$ contribution from multiple scattering, varying
single impurity strength $\bar{U}\equiv \left(\textrm {imp. volume}\right) \times \left(\textrm {imp. potential}\right)$ and
disorder strength ($n_{i}\bar{U}^2$)
for
fixed filling factor $\beta=n_{i} \times \left(\textrm{imp. volume}\right)$ and
chemical potential. The black solid line indicates the boundary where
$\sigma_{xy}^{\rm{ext,1}}=\sigma_{xy}^{\rm{int}}(\mu=0)=2\frac{e^{2}}{\hbar}\frac{\sqrt{2m\lambda}}{4\pi^{2}}$.
The plot is bounded from above due to the restriction of the impurities being point-like, $\epsilon_{F}a\ll1$
(here we take $\epsilon_{F}a<0.1$). (b-c) $\sigma_{xy}^{\rm{int}},\sigma_{xy}^{\rm{ext,1}}$ (b)
and total $\rho_{xy}$ (c) for varying chemical potential at two points
of the disorder parameter space in marked by (i) and (ii) in the upper panel.
$\rho_{xy}$ is normalized by $\rho_{xx}^{2}\sigma_{xy}^{\rm{int}}$ at
$\mu=0$ of point (i), $\sigma_{xy}$ is given in units of $\frac{e^{2}}{\hbar}\frac{\sqrt{2m\lambda}}{4\pi^{2}}$.
\label{fig:sigma_xy multiple scattering colorbar}}
\end{figure}

In Fig. \ref{fig:sigma_xy multiple scattering colorbar}, $\sigma_{xy}^{\rm ext,1}$ is plotted for varying impurity strength and concentration at a given value of the chemical potential. We identify two regimes in the disorder parameter space, separated by the black line which marks the boundary $\sigma_{xy}^{\rm int}=\sigma_{xy}^{\rm ext,1}$. Above the black line, the skew-scattering mechanism dominates. Below the black line, the impurities may or may not be at resonance, where the scattering phases are enhanced. The condition for resonance requires fine-tuning between the height of the impurity potential and its size (see Appendix \ref{sec:scat-phase-appendix}). Therefore, the resonant regions occupy a small fraction of the parameter space, corresponding to narrow regions of large $\left| \sigma_{xy}^{\rm ext,1} \right|$ in Fig. \ref{fig:sigma_xy multiple scattering colorbar}. Away from these regions, the conductivity is dominated by the intrinsic mechanism.

Excluding the resonant regions, the skew-scattering term increases with the density of states.
Therefore, the skew-dominated region expands as the chemical potential is increased (Fig. \ref{fig:sigma_xy multiple scattering colorbar}).




An analysis of the asymptotics of $\rho_{xy}$  leads to the temperature dependence $\rho_{xy}(T)-\rho_{xy}(0)\sim T^2$ for $T \ll \lambda$ and
 $\rho_{xy}(T) \sim T^{3/2}$ for $T \gg \lambda$ (see Supplemental Material).




\section{Conclusions}

To conclude, we have studied the anomalous Hall effect in a TRS-broken Weyl semimetal
with a single pair of Weyl nodes and rotational symmetry around one axis.
The anomalous Hall conductivity results from both the intrinsic and extrinsic mechanisms.
The former is disorder independent and dictated only by the band structure, 
while the latter comes from electron-disorder scattering.
The intrinsic part of the anomalous Hall conductivity slowly varies as a function of the chemical potential, from the pseudo-quantized value  $\frac{e^2}{2 \pi h}\Delta_k$ at the Weyl node to
$0.94 \frac{e^2}{2 \pi h}\Delta_k$ at the Lifshitz transition point.

The extrinsic part of the anomalous Hall conductivity dominates in the clean 
limit for a chemical potential at a finite distance away from the Weyl nodes. We have focused on the case of small size impurities.
In this case, the elastic scattering of the electrons is described by two scattering phases $\delta_{\pm}$, 
corresponding to the scattering eigenstates with total angular momentum projection on z-axis $j_z=\pm 1/2$. 
We have computed the phases $\delta_{\pm}$ as a function of the impurity parameters.
The scattering rates can then be expressed in terms of these phase shifts.
Then, one can estimate the magnitude of the skew-scattering and side-jump processes, and determine the disorder regime in which the skew-scattering mechanism dominates. This is the case for dilute and either strong or resonant impurities. We focus on this disorder regime and derive analytic results for the anomalous Hall transport coefficients.  We analyse their asymptotic behaviour as a function of the chemical potential and temperature.
We find that at low temperatures, the Hall resistivity increases quadratically with temperature, while at high temperatures it scales as $T^{3/2}$.




While our work focused on Weyl semimetals, we expect the results
to hold for a variety of range of materials with two bands and spin-orbit interaction.

\begin{acknowledgments}
The authors are grateful to  B. Yan, T. Holder and D. Kaplan for useful discussions.
This research was supported by ISF-China 3119/19 and ISF 1355/20. Y. M. thanks the PhD scholarship of the Israeli Scholarship
Education Foundation (ISEF) for excellence in academic and social leadership.
\end{acknowledgments}

\appendix

\section{Scattering phases \label{sec:scat-phase-appendix}}

In the case of strong impurities, one has to calculate the scattering
T-matrix to get the scattering rate. The theory for calculating the
T-matrix in anisotropic systems for low-energy electrons will be presented
in a separate paper, and here we
provide a brief summary of the theory and its application to our Weyl-semimetal
model.

We consider impurities of a very small size compared to the Fermi
wavelength. First we treat the spinless case and later generalize.
In this case, the impurity scattering will be dominated by a single
scattering channel, which is analagous to s-wave channel in an isotropic
system.

The scattered eigenstate of the s-wave-like channel is a superposition
of incoming and outgoing waves with a phase shift caused by the impurity:

\begin{equation}
\psi_{s}^\textrm{scat.}=\psi_{s}^\textrm{in}+e^{2i\delta_{s}}\psi_{s}^\textrm{out}.\label{eq:scattering state}
\end{equation}
In the limit of a small impurity, the outgoing (incoming) states
corresponding to this channel are given by

\begin{equation}
\psi_{s}^\textrm{out(in)}(\mathbf{r})\simeq G_{\epsilon}^{R(A)}(\mathbf{r},0),\label{eq:scattering states}
\end{equation}
with $G^{R(A)}$ being the retarded (advanced) Green function of the
clean Hamiltonian, $G_{\epsilon}^{R(A)}=\frac{1}{\epsilon-H\pm i\delta}$.

The Lippmann-Schwinger equation leads for the following formula for
the T-matrix:

\begin{equation}
T_{\epsilon}=V\left[1-G_{\epsilon}^{R}V\right]^{-1}.
\end{equation}
Let us define the space-averaged T-matrix,

\begin{equation}
\bar{T}_{\epsilon}=\intop d^{3}r\intop d^{3}r'T_{\epsilon}(\mathbf{r},\mathbf{r}').\label{eq:space avg T-matrix}
\end{equation}
It is related to the scattering matrix of the s-wave-like channel
by

\begin{equation}
\bar{T}_{\epsilon}=\frac{1-S}{-\left(G_{\epsilon}^{R}(\mathbf{r}-\mathbf{r}'=0)-G_{\epsilon}^{A}(\mathbf{r}-\mathbf{r}'=0)\right)}=\frac{1-S}{2\pi i\nu(\epsilon)},\label{eq:T bar definition}
\end{equation}
where $S$ is the scattering matrix $S=e^{i2\delta_{s}}$ and $\nu(\epsilon)$
is the density of states at the appropriate energy.

To calculate the phases, we define the R-matrix by

\begin{equation}
R_{\epsilon}=\frac{T_{\epsilon}+T_{\epsilon}^{\dagger}}{2}=V\left[1-\frac{G_{\epsilon}^{R}+G_{\epsilon}^{A}}{2}V\right]^{-1}.\label{eq:R matrix def}
\end{equation}
The space-averaged R-matrix is defined in the same manner as
for the T-matrix:
\begin{equation}
\bar{R}_{\epsilon}=\intop d^{3}r\intop d^{3}r'R_{\epsilon}(\mathbf{r}',\mathbf{r}'').
\end{equation}
It is related to the main scattering phases by

\begin{equation}
\tan\left(\delta(\epsilon)\right)=-\pi\bar{R}_{\epsilon}\nu(\epsilon).\label{eq:r matrix and phases}
\end{equation}
At the low-energy limit we can treat $\bar{R}$ as energy-independent.
We will solve it directly by diagonalizing Eq. (\ref{eq:R matrix def}).

Generalizing to spinful electrons, the objects defined above now have
a matrix structure in spin space, and there will be two primary scattering
eigenstates and phases. The main phases are then given by the generalization
of Eq. (\ref{eq:r matrix and phases}),

\begin{equation}
\tan(\delta_{\pm}(\epsilon))=-\pi\bar{R}_{\pm}\nu_{\pm}(\epsilon),\label{eq:r matrix and phases spinful}
\end{equation}
where $\bar{R}_{\pm}$ are the eigenvalues of $\bar{R}$, and $\nu_{\pm}(\epsilon)$
is the density of states in the projected subspaces of the corresponding
eigenvectors.

For systems and impurities with rotation symmetry around one axis,
$\bar{R}$ will be diagonal in the appropriate basis (e.g. eigenstates
of $J_{z}=L_{z}+\frac{\sigma_{z}}{2}$ for rotation symmetry around
z). In this basis, the space-averaged T-matrix is written as

\begin{widetext}
\begin{equation}
\bar{T}(\epsilon)=\frac{1}{2\pi i}\begin{bmatrix}\frac{1-S_{-}}{\nu_{+}} & 0\\
0 & \frac{1-S_{-}}{\nu_{-}}
\end{bmatrix}=\frac{1}{2\pi i}\begin{bmatrix}\frac{1-\exp\left(2i\delta_{+}\right)}{\nu_{+}} & 0\\
0 & \frac{1-\exp\left(2i\delta_{-}\right)}{\nu_{-}}
\end{bmatrix}\equiv\bar{T}_{0}+\bar{T}_{z}\sigma_{z},
\label{eq:T_bar expression}
\end{equation}

where we denoted

\begin{align}
\bar{T}_{0}(\epsilon) & =\frac{1}{4\pi i}\left(\frac{1-\exp\left(2i\delta_{+}\right)}{\nu_{+}}+\frac{1-\exp\left(2i\delta_{-}\right)}{\nu_{-}}\right), \label{eq:T matrix components WSM T0}\\
\bar{T}_{z}(\epsilon) & =\frac{1}{4\pi i}\left(\frac{1-\exp\left(2i\delta_{+}\right)}{\nu_{+}}-\frac{1-\exp\left(2i\delta_{-}\right)}{\nu_{-}}\right).
\label{eq:T matrix components WSM Tz}
\end{align}

\subsection{Scattering phases in a Weyl semimetal}

Let us consider the model discussed in the paper for TRS-breaking
WSM:

\begin{equation}
H=u_{\parallel}\left(p_{x}\sigma_{x}+p_{y}\sigma_{y}\right)+\left(\frac{p_{z}^{2}}{2m}-\lambda\right)\sigma_{z}+V(\mathbf{r}),\qquad\qquad V(\mathbf{r})=\begin{cases}
U & \left|x^{2}+y^{2}\right|<a,\left|z\right|<b,\\
0 & \rm{else}.
\end{cases}
\end{equation}

For convenience, in this section we will set the xy-plane Fermi velocity
to one, $u_{\parallel}=1$, and it should be restored when necessary
for correct dimensions. In the absence of full rotation symmetry,
there is no straight-forward analytical expression for the incoming
and outgoing eigenfunctions. Therefore, we numerically solve Eq. (\ref{eq:R matrix def})
in the limit $ka\ll1$. Considering impurities symmetric under rotations
around the z-axis, $j_{z}$ is a good quantum number for the scattering
eigenstates, since

\begin{equation}
\left[H,J_{z}\right]=0,
\end{equation}
with $J_{z}\equiv L_{z}+\frac{\sigma_{z}}{2}$. We have considered
a cylindrical impurity to simplify the calculation, but we expect
the results for spherical impurities to give the same qualitative
results. We define a dimensionless aspect ratio for the cylindrical
impurity

\begin{equation}
F=\frac{\pi a}{2mb^{2}}.
\end{equation}

In Fig. S\ref{fig:numerical T WSM} we display the results with the dimensionless parameters
$\tilde{U}\equiv\frac{Ua}{\pi}$, $\tilde{R}_{\pm}=\frac{\bar{R}_{\pm}}{2\pi^{2}ab}$,
for different aspect ratio values. Generally, we see resonant behavior
at values of the order $\tilde{U}=1$ (modulu one). In the limit $\tilde{U}\ll1$,
the Born approximation result is restored, $\tilde{R}_{\pm}\simeq\tilde{U}$.
For values $\tilde{U}\gtrsim1$, the Born approximation is meaningless
and the exact values of $\tilde{R}_{\pm}$ have to be taken from the
numerics, since the resonances appear to be quasi-periodic. The resonances
of the two diagonal elements do not occur together, so for $\tilde{U}\gtrsim1$
one can expect a finite difference $\tilde{R}_{+}-\tilde{R}_{-}$ of order
unity, leading to a phase difference calculated by Eq. (\ref{eq:r matrix and phases spinful}).

\begin{figure}[H]
	\centering
	
	\includegraphics[scale=0.5]{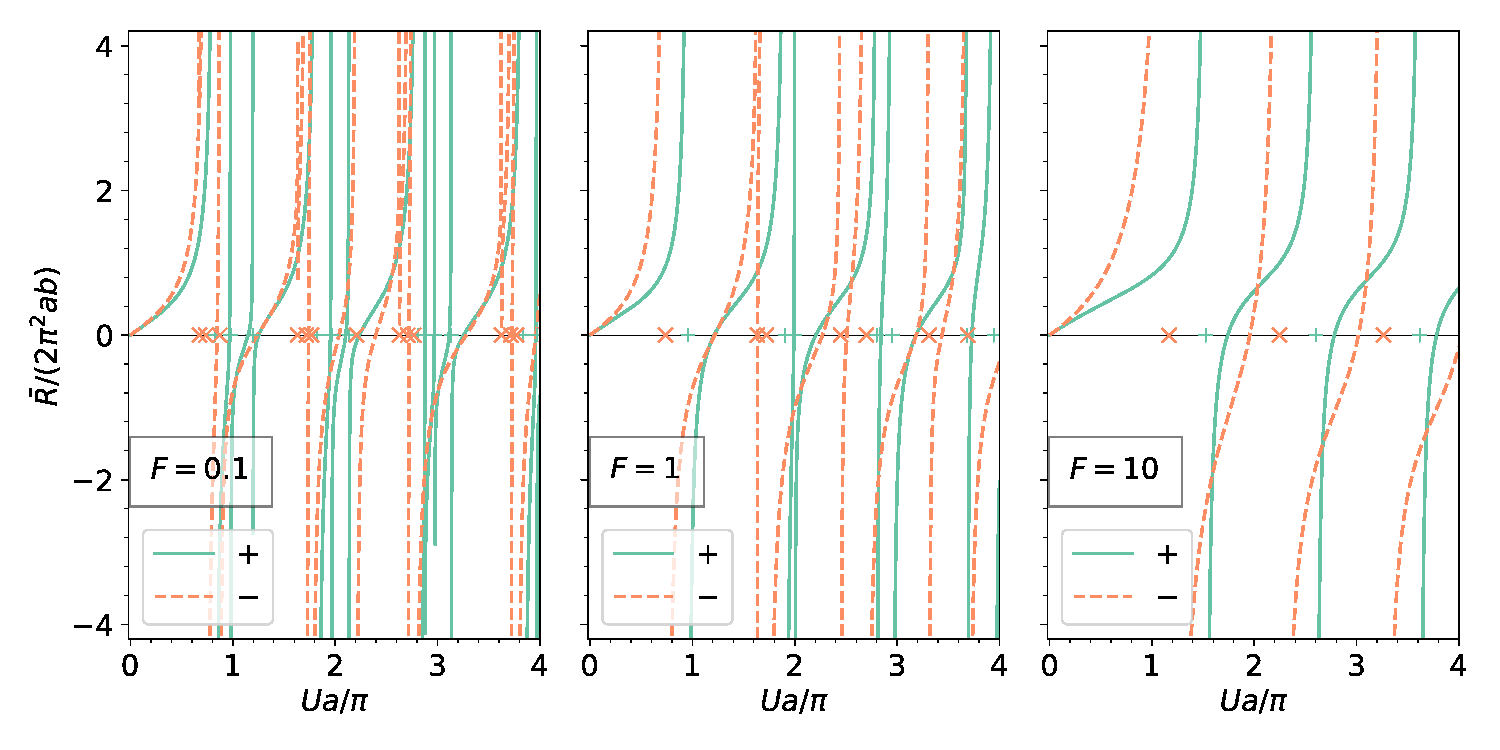}
	
	\caption{Numerical calculation of $\bar{R}_{\pm}$ for cylindrical impurity
		with different aspect ratios.\label{fig:numerical T WSM}}
\end{figure}

\section{Anomalous Hall conductivity within Boltzmann equation approach}

Here we give the technical details of solving the Boltzmann equation
to find the linear response to an electric field.

\subsection{Spectrum and plane-wave eigenbasis}

The Hamiltonian [Eq. (1) in the main text] has two bands with energies $\epsilon_{e/h,\mathbf{k}}=\pm\sqrt{u_{\parallel}^{2}\left(k_{x}^{2}+k_{y}^{2}\right)+\left(\frac{k_{z}^{2}}{2m}-\lambda\right)^{2}}$.
The plane-wave eigenstates are $\psi_{e/h,\mathbf{k}}(\mathbf{r})=e^{i\mathbf{k}\mathbf{r}}u_{e/h,\mathbf{k}}$,
with $u_{e/h,\mathbf{k}}$ being the eigenspinors

\begin{equation}
u_{e\mathbf{k}}=\begin{pmatrix}\cos\frac{\theta}{2}\\
\sin\frac{\theta}{2}e^{i\varphi}
\end{pmatrix},\qquad\qquad u_{h\mathbf{k}}=\begin{pmatrix}-\sin\frac{\theta}{2}\\
\cos\frac{\theta}{2}e^{i\varphi}
\end{pmatrix},
\end{equation}
where the direction of the spinor is given by $\cos\theta=M/\epsilon_{\mathbf{k}}=\left(k_{z}^{2}/(2m)-\lambda\right)/\epsilon_{\mathbf{k}}$,
$\tan\varphi=k_{y}/k_{x}$. The density of states is given by
\begin{equation}
\nu(\epsilon)=\sum_{b=e,h}\intop\left(d^{3}k\right)\delta(\epsilon-\epsilon_{b\mathbf{k}})   =\frac{\left|\epsilon\right|\sqrt{2m\lambda}}{2\pi^{2}}\begin{cases}
\sqrt{1+\frac{\left|\epsilon\right|}{\lambda}}-\sqrt{1-\frac{\left|\epsilon\right|}{\lambda}} & \left|\epsilon\right|<\lambda,\\
\sqrt{1+\frac{\left|\epsilon\right|}{\lambda}} & \left|\epsilon\right|>\lambda.
\end{cases}
\end{equation}
We define the densities of states for the projection on the $s_{z}=\pm1/2$
pseudospins by

\begin{equation}
\nu_{\pm}(\epsilon)=\sum_{b=e,h}\intop\left(d^{3}k\right)\frac{1\pm \cos\theta}{2}\delta(\epsilon-\epsilon_{b\mathbf{k}})=\frac{\nu(\epsilon)}{2}\left(1\pm B(\epsilon)\right),
\end{equation}
where 
\begin{equation}
B(\epsilon)\equiv\frac{1}{\nu(\epsilon)}\sum_{b=e,h}\intop\left(d^{3}k\right)\cos\theta\delta(\epsilon-\epsilon_{b\mathbf{k}}) = -\frac{\lambda}{3\epsilon}\begin{cases}
1-\sqrt{1-\frac{\epsilon^{2}}{\lambda^{2}}} & \left|\epsilon\right|<\lambda,\\
2-\frac{\left|\epsilon\right|}{\lambda} & \left|\epsilon\right|>\lambda
\end{cases}
\end{equation}
gives the relative pseudospin polarization at energy $\epsilon$.

\subsection{Scattering rates}

The scattering rate is calculated by the
Fermi golden rule for the single impurity T-matrix:

\begin{equation}
w_{b'\mathbf{k}',b\mathbf{k}}=2\pi\delta\left(\epsilon_{b'\mathbf{k}'}-\epsilon_{b\mathbf{k}}\right)n_{i}\left|T_{b'\mathbf{k}',b\mathbf{k}}\right|^{2}.
\end{equation}
Here, $T$ is the retarded T-matrix of a single impurity. For
point-like impurities, the T-matrix element between states depends
on the momentum only through the spinor structure:

\begin{equation}
T_{b\mathbf{k},b'\mathbf{k}'}=\left\langle u_{b\mathbf{k}}\left|\bar{T}\right|u_{b'\mathbf{k}'}\right\rangle,
\end{equation}
where $\bar{T}$ is given by Eq. (\ref{eq:T_bar expression}). We expand the projector to the $u_{b\mathbf{k}}$
eigenstate:

\begin{align}
\left|u_{b\mathbf{k}}\right\rangle \left\langle u_{b\mathbf{k}}\right| & =\sum_{\alpha=0,xyz}P_{\alpha}^{b}(\hat{k})\frac{\sigma_{\alpha}}{2},
\end{align}
with

\begin{align}
P_{0}^{e/h}(\hat{k})  =1, \qquad P_{x}^{e/h}(\hat{k})  = \pm \sin\theta\cos\varphi, \nonumber \\
P_{y}^{e/h}(\hat{k})  =\pm \sin\theta\sin\varphi, \qquad P_{z}^{e/h}(\hat{k}) =\pm \cos\theta.
\end{align}
This allows us to write the scattering amplitudes between two states
as products of functions of a single momentum coordinate (since the scattering is elastic, we suppress the band indices from hereon):

\begin{align}
\left.W_{\mathbf{k},\mathbf{k}'}\right|_{(\epsilon_{\mathbf{k}}=\epsilon_{\mathbf{k}'})}\equiv2\pi n_{i}\left|T_{\mathbf{k},\mathbf{k}'}\right|_{(\epsilon_{\mathbf{k}}=\epsilon_{\mathbf{k}'})}^{2}  =\sum_{\alpha\beta}P_{\alpha}(\hat{k})P_{\beta}(\hat{k}')W_{\alpha\beta},\label{eq:scattering amplitude}
\end{align}
where we defined 
\begin{equation}
W_{\alpha\beta}\equiv\frac{1}{2}\pi n_{i}\textrm{Tr}\left\{ \sigma_{\alpha}\bar{T}\sigma_{\beta}\bar{T}^{\dagger}\right\}. \label{eq:W alpha beta}
\end{equation}
In our system, the matrix $\bar{T}$ has only $\sigma_0, \sigma_z$ components [Eqs. (\ref{eq:T matrix components WSM T0}), (\ref{eq:T matrix components WSM Tz})].
The resulting non-zero elements of $W_{\alpha\beta}$ are:

\begin{align}
W_{00} & =W_{zz}=\pi n_{i}\left(\left|\bar{T}_{0}\right|^{2}+\left|\bar{T}_{z}\right|^{2}\right)=\frac{n_{i}}{2\pi}\left[\frac{\sin^{2}\delta_{+}}{\nu_{+}^{2}}+\frac{\sin^{2}\delta_{-}}{\nu_{-}^{2}}\right]\label{eq:W_alphabeta first},\\
W_{0z} & =W_{z0}=2\pi n_{i}{\rm Re}\left(\bar{T}_{0}\bar{T}_{z}^{\ast}\right)=\frac{n_{i}}{2\pi}\left[\frac{\sin^{2}\delta_{+}}{\nu_{+}^{2}}-\frac{\sin^{2}\delta_{-}}{\nu_{-}^{2}}\right],\\
W_{xx} & =W_{yy}=\pi n_{i}\left(\left|\bar{T}_{0}\right|^{2}-\left|\bar{T}_{z}\right|^{2}\right)=\frac{n_{i}\sin\delta_{+}\sin\delta_{-}\cos\left(\delta_{+}-\delta_{-}\right)}{\pi\nu_{+}\nu_{-}},\\
W_{xy} & =-W_{yx}=-2\pi n_{i}{\rm Im}\left(\bar{T}_{0}\bar{T}_{z}^{\ast}\right)=\frac{n_{i}\sin\delta_{+}\sin\delta_{-}\sin\left(\delta_{+}-\delta_{-}\right)}{\pi\nu_{+}\nu_{-}}.\label{eq:W_alphabeta last}
\end{align}

\subsubsection{Linear response distribution function}

Here we present the solution of the Boltzmann equation for the distribution
function. First let us solve generally for spinors, utilizing the
form of the scattering amplitude given in Eq. (\ref{eq:scattering amplitude}).
We write the linear correction in the form

\begin{equation}
\delta f(\mathbf{k})=-\frac{\partial f_{0}(\epsilon_{\mathbf{k}})}{\partial\epsilon_{\mathbf{k}}}\chi(\hat{k}),
\end{equation}
where $\hat{k}$ is a vector parameterizing the Fermi surface. Then,
the Boltzmann equation is solved for any given energy $\epsilon_{\mathbf{k}}=\epsilon$,
which will be implicit from hereon. Keeping the terms linear in $\overrightarrow{E}$,
we obtain

\begin{equation}
e\overrightarrow{E}\cdot\overrightarrow{v}(\hat{k})=\intop\left(d\mathbf{k}'\right)\delta(\epsilon-\epsilon_{\mathbf{k}'})W(\hat{k},\hat{k}')\left(\chi(\hat{k})-\chi(\hat{k}')\right).\label{eq:boltzmann eq linear order}
\end{equation}
We will absorb the density of states in the integration measure by
defining $d\left[\hat{k}\right]\equiv\left(d\mathbf{k}\right)\delta(\epsilon-\epsilon_{\mathbf{k}})$,
so that $\intop\left(d\mathbf{k}\right)\delta(\epsilon-\epsilon_{\mathbf{k}})=\intop d\left[\hat{k}\right]=\nu(\epsilon)$.
The collision integral in Eq. (\ref{eq:boltzmann eq linear order})
can be written as

\begin{equation}
\intop d\left[\hat{k}'\right]W(\hat{k},\hat{k}')\left(\chi(\hat{k})-\chi(\hat{k}')\right)=\frac{\chi(\hat{k})}{\tau_{\mathbf{k}}^{{\rm el}}}-\sum_{\alpha\beta}W_{\alpha\beta}P_{\alpha}(\hat{k})\intop d\left[\hat{k}'\right]P_{\beta}(\hat{k}')\chi(\hat{k}'),
\end{equation}
where 
\begin{equation}
1/\tau_{\mathbf{k}}^{{\rm el}} \equiv \intop d\left[\hat{k'}\right]W(\hat{k},\hat{k}') \label{eq:elastic rate def}
\end{equation}
is the elastic scattering rate. To solve the integral equation, let
us denote

\begin{equation}
\intop d\left[\hat{k}'\right]P_{\beta}(\hat{k}')\chi(\hat{k}')=x_{\beta}.\label{eq:k_prime_integral}
\end{equation}
We obtain

\begin{equation}
\chi(\hat{k})=\tau_{\mathbf{k}}^{{\rm el}}\left(e\overrightarrow{E}\cdot\overrightarrow{v}(\hat{k})+\sum_{\alpha\beta}W_{\alpha\beta}x_{\beta}P_{\alpha}(\hat{k})\right).\label{dist correction}
\end{equation}
We have obtained the correction to the distribution as a finite sum
of known functions. It is left to determine the coefficients $x_{\alpha}$
by inserting the last equation into Eq. (\ref{eq:k_prime_integral}),
leading to

\begin{equation}
x_{\alpha} =\underbrace{\intop d\left[\hat{k}'\right]P_{\alpha}(\hat{k}')\tau_{\mathbf{k}'}^{{\rm el}}e\overrightarrow{E}\cdot\overrightarrow{v}(\hat{k}')}_{\equiv m_{\alpha}}+\sum_{\gamma}x_{\gamma}\underbrace{\sum_{\beta}W_{\beta\gamma}\intop d\left[\hat{k}'\right]\tau_{\mathbf{k}'}^{{\rm el}}P_{\beta}(\hat{k}')P_{\alpha}(\hat{k}')}_{\equiv Q_{\alpha\gamma}},
\end{equation}
resulting in a simple matrix equation

\begin{equation}
\left(\delta_{\alpha\gamma}-Q_{\alpha\gamma}\right)x_{\gamma} =m_{\alpha}.\label{eq:x gamma solution and def of m and Q}
\end{equation}
At this point, one would invert the matrix $1-Q$ in order to find
the coefficients $x_{\alpha}$ and determine the distribution
function. However, some care is needed, since this matrix has at least
one null eigenvector corresponding to the zero modes of the collision
integral. Indeed, for any constant $\chi(\hat{k})=\chi$ it is immediate
that $\intop d\left[\hat{k}'\right]\left(\chi(\hat{k})-\chi(\hat{k}')\right)=0$.
Therefore, to be able to solve this equation, $m_\alpha$ must be in the span of $1-Q$. If it is so, one can project Eq. (\ref{eq:x gamma solution and def of m and Q})
to the subspace orthogonal to the null eigenspace of $1-Q$ and proceed.

Let us now focus on our WSM model which possesses rotational symmetry
in the x-y plane, simplifying the problem. The matrix $1-Q$ is block
diagonal with two blocks: $0,z$ and $x,y$ [Eqs. (\ref{eq:W_alphabeta first})-(\ref{eq:W_alphabeta last})].
For an electric field in the x-y plane, the non-zero elements of $m_{\alpha}$
are only $\alpha=x,y$. There are no zero modes for $1-Q$ in the
x-y plane. Thus, we can project to this plane and safely invert:

\begin{equation}
x_{\alpha}=\left(1-Q\right)_{\alpha\beta}^{-1}m_{\beta},
\end{equation}
where $\alpha,\beta\in\left\{ x,y\right\} $. Due to the rotational
symmetry, it is convenient to define 
\begin{equation}
P_{\pm}(\hat{n})=\frac{1}{\sqrt{2}}\left(P_{x}(\hat{n})\pm iP_{y}(\hat{n})\right),
\end{equation}
and similarly for $E_{\pm},v_{\pm}(\hat{n}),j_{\pm}$ and $m_{\pm}$.
In this basis, the $Q$ matrix is diagonal (transforming the basis
by calculating the $W_{\alpha\beta},Q_{\alpha\beta}$ matrix elements
from Eqs. (\ref{eq:W alpha beta}), (\ref{eq:x gamma solution and def of m and Q})
for components $\alpha,\beta\in\left\{ +,-\right\} $),

\begin{equation}
Q_{\left(\alpha\beta\in\left\{ +,-\right\} \right)}=\nu(\epsilon)\left\langle \frac{\tau_{\mathbf{k}}^{{\rm el}}\sin^{2}\theta}{2}\right\rangle \begin{bmatrix}W_{-+} & 0\\
0 & W_{+-}
\end{bmatrix},
\end{equation}
where the angular brackets indicate averaging over the Fermi surface,
\begin{equation}
\langle g(\mathbf{k})\rangle \equiv \int d \left[ \hat{k} \right] g(\mathbf{k})/\nu,
\end{equation}
and the scattering elements in the new basis are
\begin{equation}
W_{\pm\mp}=W_{xx}\pm iW_{xy}=\frac{n_{i}\sin\delta_{+}\sin\delta_{-}}{\pi\nu_{+}\nu_{-}}\exp\left(i(\delta_{+}\pm\delta_{-})\right).
\end{equation}
The inversion of the matrix $1-Q$ is now straightforward. Setting
$E_{-}=0$ we obtain $x_{-}=m_{-}=0$ and

\begin{align}
m_{+} & =eE_{+}\nu(\epsilon)u_{\parallel}^{2}\left\langle \frac{\tau_{\mathbf{k}}^{{\rm el}}\sin^{2}\theta}{2}\right\rangle, \\
x_{+} & =\frac{1}{1-Q_{++}}m_{+},
\end{align}
giving the solution for the distribution function correction

\begin{equation}
\chi(\hat{k})=\tau_{\mathbf{k}}^{{\rm el}}\left(eE_{+}v_{-}(\hat{k})+P_{-}(\hat{k})W_{-+}x_{+}\right).
\end{equation}
Due to the linear dispersion in x-y plane, $v_{x,y}(\hat{k})=u_{\parallel}P_{x,y}(\hat{k})$.
Noting that

\begin{equation}
\frac{u_{\parallel}^{2}\sin^{2}\theta}{2}=\intop\frac{d\varphi}{2\pi}v_{x}^{2}=\intop\frac{d\varphi}{2\pi}v_{y}^{2},
\end{equation}
we replace $u_{\parallel}^{2}\sin^{2}\theta/2\rightarrow v_{\mathbf{k},\parallel}^{2}$,
where $v_{\mathbf{k},\parallel}^{2}$ is the velocity in any
direction along the x-y plane. We denote the real and imaginary parts of the denominator $1-Q_{++}$ by 

\begin{align}
1/r^{\parallel} & \equiv 1-\nu(\epsilon)\left\langle \frac{\tau_{\mathbf{k}}^{{\rm el}} v_{\mathbf{k},\parallel}^{2}}{u_{\parallel}^{2}}\right\rangle {\rm Re}(W_{-+})=1-\frac{n_{i}\nu\sin\delta_{+}\sin\delta_{-}\cos\left(\delta_{+}-\delta_{-}\right)}{\pi u_{\parallel}^{2}\nu_{+}\nu_{-}}\left\langle \tau_{\mathbf{k}}^{{\rm el}} v_{\mathbf{k},\parallel}^{2}\right\rangle, \\
1/r^{\perp} & \equiv -\nu(\epsilon)\left\langle \frac{\tau_{\mathbf{k}}^{{\rm el}} v_{\mathbf{k},\parallel}^{2}}{u_{\parallel}^{2}}\right\rangle {\rm Im}(W_{-+})=\frac{n_{i}\nu\sin\delta_{+}\sin\delta_{-}\sin\left(\delta_{+}-\delta_{-}\right)}{\pi u_{\parallel}^{2}\nu_{+}\nu_{-}}\left\langle \tau_{\mathbf{k}}^{{\rm el}} v_{\mathbf{k},\parallel}^{2}\right\rangle.
\end{align}
Then we have

\begin{equation}
\chi(\hat{k})=eE_{+}v_{-}(\hat{k})\frac{1/\tau^{\parallel}_{\mathbf{k}}-i/\tau^{\perp}_{\mathbf{k}}} {1/{\left( \tau^{\parallel}_{\mathbf{k}}\right) }^{2}+1/{\left( \tau^{\perp}_{\mathbf{k}} \right)}^{2}},
\end{equation}
where we defined
\begin{align}
\tau^{\parallel}_{\mathbf{k}} & =r^{\parallel}\tau^{{\rm el}}_{\mathbf{k}},\\
\tau^{\perp}_{\mathbf{k}} & =r^{\perp}\tau^{{\rm el}}_{\mathbf{k}}.
\end{align}
Finally we explicitly calculate the elastic scattering rate [Eq. (\ref{eq:elastic rate def})],
\begin{equation}
1/\tau_{\mathbf{k}}^{{\rm el}}=\frac{n_{i}}{\pi}\left[\frac{\sin^{2}\delta_{+}}{\nu_{+}} + \frac{\sin^{2}\delta_{-}}{\nu_{-}} + \left( \frac{\sin^{2}\delta_{+}}{\nu_{+}} - \frac{\sin^{2}\delta_{-}}{\nu_{-}} \right) \cos\theta \right],
\end{equation}
as well as the integral corresponding to the vertex correction:

\begin{equation}
\left\langle \tau_{\mathbf{k}}^{{\rm el}}v_{\mathbf{k},\parallel}^{2}\right\rangle   =\frac{\epsilon\sqrt{2m\lambda}}{4\pi^{2}\nu W_{00}}I(\epsilon,\frac{W_{00}}{W_{0z}}),
\end{equation}
where we defined
\begin{align}
I(\epsilon,x) & \equiv\begin{cases}
\frac{2x\left(3x+2\frac{\lambda}{\epsilon}\right)}{3}\left(\sqrt{1+\frac{\left|\epsilon\right|}{\lambda}}-\sqrt{1-\frac{\left|\epsilon\right|}{\lambda}}\right)-{\rm sgn}(\epsilon)\frac{2x}{3}\left(\sqrt{1+\frac{\left|\epsilon\right|}{\lambda}}+\sqrt{1-\frac{\left|\epsilon\right|}{\lambda}}\right)\\
+\frac{2x\left(x^{2}-1\right)\epsilon}{\sqrt{\lambda\left|\lambda-x\epsilon\right|}}\left[F(\frac{\lambda+\left|\epsilon\right|}{\lambda-x\epsilon})-F(\frac{\lambda-\left|\epsilon\right|}{\lambda-x\epsilon})\right] & \left|\epsilon\right|<\lambda,\\
\frac{2x\left(3x+2\frac{\lambda}{\epsilon}-{\rm sgn}(\epsilon)\right)}{3}\sqrt{1+\frac{\left|\epsilon\right|}{\lambda}}+\frac{2x\left(x^{2}-1\right)\epsilon}{\sqrt{\lambda\left|\lambda-x\epsilon\right|}}F(\frac{\lambda+\left|\epsilon\right|}{\lambda-x\epsilon}) & \left|\epsilon\right|>\lambda,
\end{cases}\\
F(y) & \equiv\begin{cases}
{\rm arctanh}\left[\min\left(\sqrt{y},\sqrt{\frac{1}{y}}\right)\right] & y>0,\\
-\arctan\left(\sqrt{-y}\right) & y<0.
\end{cases}
\end{align}

\end{widetext}

\subsubsection{Longitudinal and anomalous Hall conductivities}

Having solved the distribution function, we now proceed to calculate
the conductivities. After calculating the expectation value of $j_+=j_x+i j_y$ we extract both
the longitudinal and Hall conductivities from the relation
\begin{equation}
j_+=\left(\sigma_{xx}\mp i\sigma_{xy}\right)E_+.
\end{equation}
For the longitudinal conductivity we get the standard result,

\begin{align}
\sigma_{xx} =e^{2}\nu(\epsilon)\left\langle \tau_{\mathbf{k}}^{{\rm el}}  v_{\mathbf{k},\parallel}^{2} \right\rangle \frac{1/r^{\parallel}}{1/\left(r^{\parallel}\right)^{2}+1/\left(r^{\perp}\right)^{2}}.
\end{align}
Next we calculate the anomalous Hall conductivity and identify the
different terms:

\begin{equation}
\sigma_{xy}=\frac{e}{E_{y}}\sum_{\mathbf{k}}v_{\mathbf{k}, x} f_\mathbf{k}=\sigma_{xy}^{\rm int} + \sigma_{xy}^{{\rm ext},1}.
\end{equation}

The intrinsic contribution $\sigma_{xy}^{\rm int}$ is due to the anomalous velocity of the Fermi-sea electrons and was already calculated in the main text.
The skew-scattering term $\sigma_{xy}^{{\rm ext,1}}$ comes from the
normal part of the velocity operator and the correction
to the distribution function proportional to $1/\tau^{\perp}$.
We find

\begin{align}
\sigma_{xy}^{{\rm ext,1}} =e^{2}\nu(\epsilon)\left\langle \tau_{\mathbf{k}}^{{\rm el}} v_{\mathbf{k},\parallel}^{2} \right\rangle \frac{1/r^{\perp}}{1/\left(r^{\parallel}\right)^{2}+1/\left(r^{\perp}\right)^{2}}=\frac{r^{\parallel}}{r^{\perp}}\sigma_{xx}.
\end{align}
In the limit $\tau^{\perp}\gg\tau^{\parallel}$,

\begin{align}
\sigma_{xx} & =e^{2}\nu(\epsilon)\left\langle \tau_{\mathbf{k}}^{{\rm \parallel}} v_{\mathbf{k},\parallel}^{2} \right\rangle, \\
\sigma_{xy}^{{\rm ext,1}} & \simeq e^{2}\nu(\epsilon)\left\langle \frac{\left(\tau_{\mathbf{k}}^{\parallel}\right)^{2}}{\tau_{\mathbf{k}}^{\perp}}v_{\mathbf{k},\parallel}^{2} \right\rangle \simeq\sin(\delta_{+}-\delta_{-})\sigma_{xx}.
\end{align}
We note that the result in the main text [Eq. (8)] is still exact in the case $\tau^{\perp}\lesssim\tau^{\parallel}$, which
may occur for resonant impurities. Although the expressions for $\sigma_{xx}$, $\sigma_{xy}$ in that case involve both $\tau_{\parallel}$ and $\tau_{\perp}$, considering the AHE resistivity leads to cancellations which result in a simpler expression.

\bibliography{paper_bib2}

\end{document}